\documentclass[12pt]{article}
\usepackage[utf8]{inputenc}
\usepackage[a4paper]{geometry}
\usepackage{amsmath,amssymb,amsfonts}
\usepackage{graphicx}
\usepackage{caption}
\usepackage{subcaption}
\usepackage{bm}
\usepackage{authblk}
\usepackage{soul, xcolor}
\usepackage{amsmath,amssymb,amsfonts}
\usepackage{setspace}
\usepackage[title]{appendix}
\usepackage{enumitem}
\usepackage[square,numbers,compress]{natbib}
\pdfoutput=1

\title{A DSMC-CFD coupling method using surrogate modelling for low-speed rarefied gas flows}

\author[1]{Giorgos Tatsios}
\author[2]{Arun K. Chinnappan}
\author[2]{Arshad Kamal}
\author[3]{Nikolaos Vasileiadis}
\author[4]{Stephanie Y. Docherty}
\author[3]{Craig White}
\author[1]{Livio Gibelli}
\author[1]{Matthew K. Borg}
\author[5]{James R. Kermode}
\author[2]{Duncan A. Lockerby}

\affil[1]{School of Engineering, Institute for Multiscale Thermofluids, University of Edinburgh, Edinburgh, EH9 3FB, United Kingdom}

\affil[2]{School of Engineering, University of Warwick, Coventry, CV4 7AL, United Kingdom}

\affil[3]{James Watt School of Engineering, University of Glasgow, Glasgow, G12 8QQ, United Kingdom}
            
\affil[4]{School of Engineering and Physical Sciences, Heriot-Watt University, Edinburgh, EH14 4AS, United Kingdom}

\affil[5]{Warwick Centre for Predictive Modelling, School of Engineering, University of Warwick, Coventry, CV4 7AL, United Kingdom}

\date{}

\begin{document}
\maketitle

\clearpage
\section*{Abstract}
A new Micro-Macro-Surrogate (MMS) hybrid method is presented that couples the Direct Simulation Monte Carlo (DSMC) method with Computational Fluid Dynamics (CFD) to simulate low-speed rarefied gas flows. The proposed MMS method incorporates surrogate modelling instead of direct coupling of DSMC data with the CFD, addressing the limitations CFD has in accurately modelling rarefied gas flows, the computational cost of DSMC for low-speed and multiscale flows, as well as the pitfalls of noise in conventional direct coupling approaches. The surrogate models, trained on the DSMC data using Bayesian inference, provide noise-free and accurate corrections to the CFD simulation enabling it to capture the non-continuum physics. The MMS hybrid approach is validated by simulating low-speed, force-driven rarefied gas flows in a canonical parallel-plate system and shows excellent agreement with DSMC benchmark results. A comparison with the typical domain decomposition DSMC-CFD hybrid method is also presented, to demonstrate the advantages of noise-avoidance in the proposed approach. The method also inherently captures the uncertainty arising from micro-model fluctuations, allowing for the quantification of noise-related uncertainty in the predictions. The proposed MMS method demonstrates the potential to enable multiscale simulations where CFD is inaccurate and DSMC is prohibitively expensive. \\

\noindent \textbf{Keywords:} Hybrid Methods; Multiscale modelling; Continuum-particle simulations; Bayesian Inference; Surrogate Modelling

\section{Introduction}
\label{sec:intro}
Understanding multiscale gas flows is critical to a number of micro technologies and low-pressure applications, ranging from photolithography machines of next-generation processor chips \cite{dgl}, thermal management systems using evaporating nanopipes \cite{nanomembrane}, and high-precision electrospray ionization mass spectrometry \cite{emerson2020}. These multiscale flows, combining rarefied gas and low-speed fluid behaviour, present a unique and formidable simulation problem. 

Conventional computational fluid dynamics (CFD) cannot capture the physics of rarefied gas flows, because it is based on the assumption of (near) local thermodynamic equilibrium --- for which there are not enough molecular collisions to attain. The departure from local quasi thermodynamic equilibrium is expressed by the Knudsen number: $\text{Kn} = \lambda/L$, where $\lambda$ is the mean free path, i.e.\ the average distance a molecule travels between molecular collisions, and $L$ is the characteristic length scale of the flow. As $\rm{Kn}$ increases (at lower pressures or in shrinking geometries), the no-slip boundary conditions and Navier-Stokes-Fourier constitutive relations (the basis of modern CFD) begin to fail, and predictions using them become unreliable. For example, measured gas flow rates in micro channels are typically a factor of two greater than those predicted by CFD~\cite{varoutis}, the drag on a micro sphere is a similar factor less~\cite{lockerby2008}, and gas molecules flow up a thermal gradient along a micro channel, instead of down~\cite{lockerby2015}. Extensions to the classic continuum model have been explored (e.g. slip boundary conditions~\cite{maxwell}, constitutive-law scaling~\cite{lockerby2008}, and higher-order governing equations~\cite{regG13,lockerby2005}), but these are only reliable for low Kn flows, and limitations include: lack of generality; impractical equation complexity; and pathological numerical instability.

The Direct Simulation Monte Carlo method, DSMC~\cite{bird}, is the state-of-the-art for solving moderate-to-high Kn flows. The DSMC method is a stochastic particle method to solve the Boltzmann equation~\cite{bird,wagner}, which converges to the Navier-Stokes equations at very low Kn. However, the accuracy of DSMC depends on simulating collisions on the scale of the mean free path, and as this decreases, the number of collisions (and number of particles) increases. Roughly speaking, to achieve the same accuracy, a flow with $\rm{Kn}=0.1$ would require a factor of 1 million times more particles than the equivalent flow with $\rm{Kn}=10$. While DSMC is really designed for high Kn flows, where it can outperform continuum models, its application to cases where there are varying densities or multiscale geometries (i.e.\ a flow field involving a range of Kn) is computationally very demanding.

Flows in micro-scale geometries typically have low-speed regions. Statistical noise is an intrinsic feature of DSMC’s stochastic nature, and reducing this to an acceptable level for low-speed flows is phenomenally expensive. A subsonic Mach 0.03 flow requires, approximately, a factor of 10,000 times more samples than a flow at Mach 3 to obtain the same signal-to-noise ratio~\cite{staterr}. For any flow with a low-speed component, the problem of noise is impenetrable. Other methods have attempted to overcome this noise barrier, including direct extensions to DSMC, but these still suffer from limitations. For example, the low-variance DSMC~\cite{lvdsmc} has not been extended to realistic gases and general cases. The Information Preservation (IP) method \cite{ip-dsmc} effectively reduces statistical scatter with limited samples; however, it entails a higher memory requirement than the regular DSMC method. Deterministic solutions of the Boltzmann equation (e.g. the Discrete Velocity Method) are noise-free, but require discretisation of the 7-dimensional phase space for unsteady 3D flows, so they suffer from prohibitive computational cost and large random-access memory requirements for the smallest of 3D geometries.

To overcome the joint challenge of physical limitations of CFD and the computational demands of particle methods, a number of hybrid strategies have been proposed. In domain-decomposition hybrid methods, the most commonly used in the literature, the domain is divided into rarefied and continuum regions. The DSMC method and conventional CFD are employed to tackle each respective region. The two methods are coupled in an overlap region, either based on state properties or gradient variables \cite{zhang,sarantis2014,boyd2008,schwartzentruberA,schwartzentruberB,halo3d,unstrhybrid}. In this hybrid approach, computational savings arise from a reduction of the required computational particles. Heterogeneous multiscale methods are not too dissimilar \cite{garcia1999,garcia2008,hmm,borg2013,docherty2014,docherty2016}, where the continuum solver is applied in the entire flow field and a number of micro patches are distributed in the domain to provide accurate local or field information to the continuum solver, such as in the form of constitutive or flux corrections. Often it is possible to decouple the DSMC mesh from the CFD one, to provide local or field corrections, which is not strictly possible with domain decomposition. A special case of heterogeneous multiscale methods is the internal-flow multiscale method (IMM)  \cite{imm,patronis2013,patronis2014,borg2015,benzi} for high-aspect-ratio internal flows, where the particle method is applied in micro-subdomains covering cross-sections of the flow domain, which are coupled based on the macroscopic conservation laws. Further computational savings are achieved in these methods from exploiting the use of asynchronous timestepping between DSMC and CFD \cite{lockerby2015}.

It is evident that hybrid multiscale methods have been developed to tackle a large range of flows, but none are suitable for low-speed 3D geometries and large ranges of Kn --- i.e.\ those required for engineering design of microscale and low-pressure systems. In conventional hybrids, the micro simulator is directly coupled to the macro simulator, via an exchanged set of fields; if the noise is above a certain level~\cite{staterr}, it can be numerically destabilising. To avoid this, data passed to the macro domain must be averaged over a large population (ensemble) of independent micro simulators (or over a long time in the case of steady-state flows). Even when the stability barrier is overcome, the hybrid results will still fluctuate in response to the underlying fluctuations of the micro model. This makes it extremely difficult to answer the crucial questions: how close is this hybrid result to the signal that we are trying to predict? How much confidence can we have in it? Working with large safety margins, as a pragmatic compromise, is not tractable for low-speed flows. The conventional hybrid approach can also be extremely wasteful and is by design unintelligent. Once the subdomains are set, the modelling input that informs the hybrid is finished. For example, after days of computer time it may turn out that the micro-model was not needed after all. A major shortcoming of the hybrid scheme is that there is no mechanism by which partial information can inform the prediction, and, in this example, stop (or switch off) the micro simulation if it is deemed redundant.

\begin{figure}
    \centering
    \includegraphics[width=0.8\textwidth]{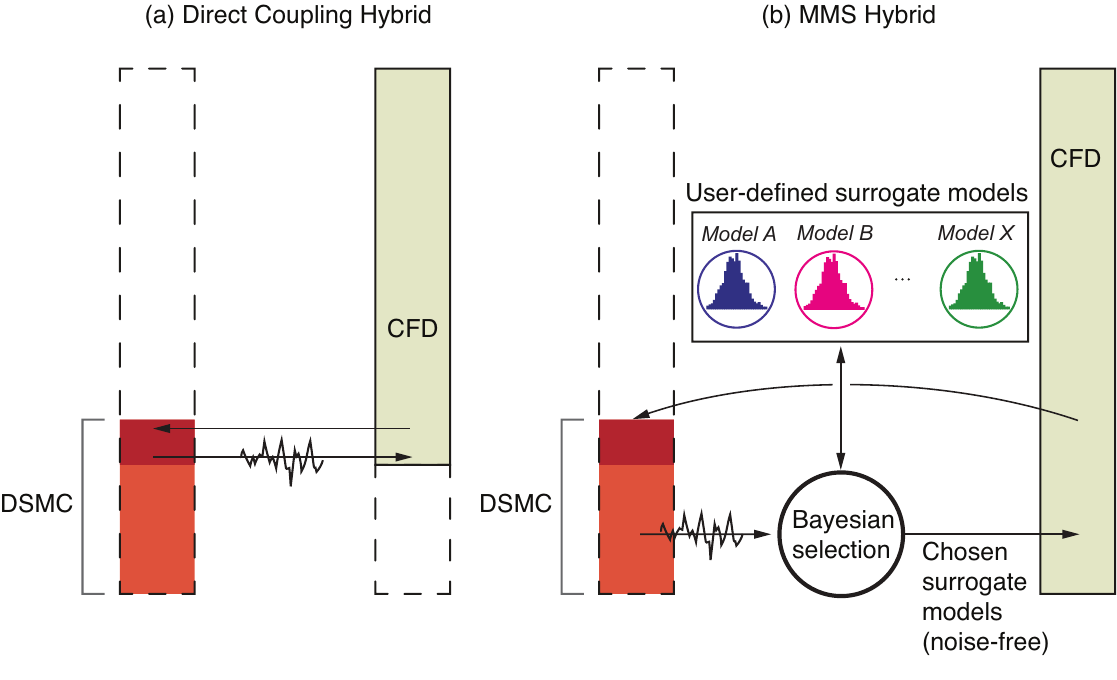}
    \caption{Comparison between (a) direct (conventional) coupling hybrid based on the widely used domain decomposition approach and (b) our proposed MMS hybrid approach. In the former approach, the noisy DSMC data directly constrains the CFD simulation, while in the latter, DSMC data is used to train noise-free surrogate models for the constitutive laws and boundary conditions (indicated by Models $A, B,\dots,X$ in the figure) that provide corrections to CFD. The arrow indicating CFD information being transferred to DSMC at the open boundary always involves the application of the Chapman-Enskog distribution defined at the local CFD quantities.}
    \label{fig1}
\end{figure}

To address these issues here we propose the Micro-Macro-Surrogate (MMS) hybrid model, as illustrated in Figure \ref{fig1}, which solves a micro model (DSMC in this case) only in regions where its accuracy is required to correct a CFD simulation (the ‘macro model’) that occupies the whole computational domain. Data measured by the micro model will no longer directly constrain the macro model. Instead, here, a new modelling component sits between the micro and macro models: the surrogate model. The surrogate is a noise-free and inexpensive substitute for the micro model, i.e.\ it is capable of capturing the non-continuum physics if it arises, either for boundary conditions or constitutive relationships. The hybrid simulation uses DSMC training data and Bayesian inference to choose (and tune) autonomously the most probable surrogate model from a hierarchy of models. The outputs of the MMS are noise-free, accurate predictions of flow fields from CFD, corrected from the surrogate model inputs, with uncertainty predictions attached to these flow fields. 

An initial setup to verify the underlying surrogate coupling strategy of the MMS is performed in a simplified case, where the DSMC component occupies the whole domain. In this case the MMS approach does not offer computational savings, but demonstrates the generality, accuracy and noise-reducing capability of the approach. In the MMS hybrid case, which is the main focus of this work, the DSMC is applied only in the Knudsen layers, to infer the relevant corrections. This strategy benefits from the computational advantages of the typical hybrid methods that save on fewer particles when the Knudsen numbers are moderate-to-low, but with less effort than a direct coupling hybrid, which suffers from the long time averaging needed to reduce the noise. The method also provides some insights into the underlying physics, through the self-selection of surrogate models, as well as assigning error bars due to the inherent probabilistic nature of the surrogate model, that have not been possible before.

The rest of the paper is organised as follows. The MMS methodology is described in Section~\ref{sec:description}. Results verifying the surrogate strategy of the MMS method are presented in Section~\ref{sec:1waycoupling}, while in Section~\ref{sec:2waycoupling} the MMS hybrid method is validated. In both cases, MMS simulations are compared with their respective benchmark DSMC results. A comparison between the direct coupling hybrid and the MMS hybrid is also presented in Section~\ref{sec:2waycoupling}. Finally, concluding remarks and future research directions are provided in Section~\ref{sec:conclusion}. 

\section{Methodology}\label{sec:description}

A Micro-Macro-Surrogate (MMS) hybrid method is presented. The microscopic model is the well-established DSMC method ~\cite{bird}, while the macroscopic model is based on conventional continuum fluid mechanics and is presented in Section~\ref{sec:macromodel}. Surrogate models are a set of probabilistic models that provide simple expressions for the boundary conditions and constitutive laws to CFD. Surrogate models replace the direct imposition of noisy DSMC data at boundaries, cell-averaged properties or face-averaged fluxes, as used by other hybrid methods. As they learn from DSMC, the surrogate models provide increasingly refined predictions, and compete with each other as the most probable choice of model to represent the current data. In the present work, surrogate models employing our physical knowledge of the system under investigation are used, instead of a generic regression. Training is performed on-the-fly as soon as DSMC data is available, and the surrogate models are only guaranteed to be accurate for the flow configuration under investigation, without aspiring to derive general governing equations \cite{datadriven}. To keep the simulation unbiased and accurate, prior information is not used, despite the resulting computational penalty. The full details of surrogate modelling are presented in Section~\ref{sec:surrogates}.

The MMS hybrid method we adopt in this work makes use of a simple iterative coupling scheme, although this can be modified in the future. The CFD model is applied in the whole flow domain, and DSMC overlays the region of the CFD where we expect it to be inaccurate, i.e. in the Knudsen layers close to the boundaries. The CFD runs first and provides data to the DSMC at the interface only (as indicated in Figure~\ref{fig1}(b)) and the incoming particles are assumed to follow the Chapman-Enskog distribution at the local CFD quantities, which has proven to be superior to the Maxwellian distribution for such hybrid simulations~\cite{hash}. The DSMC simulation runs next. The boundary and constitutive surrogate models are trained using only the DSMC data in the non-equilibrium patches. The chosen and tuned surrogate models, which represent correction fields covering the full CFD domain, are then used to correct the CFD model straightforwardly. Note that in this case a buffer region, a few cells wide, exists at the open-interface of the DSMC. DSMC data in this buffer region is not used for training. The CFD and DSMC solutions are iterated until a steady solution is reached, as is common in hybrid methods.

\subsection{Macroscopic model} \label{sec:macromodel}

In order to demonstrate the application of the method, the one-dimensional pressure-driven flow between parallel plates is considered. This simple flow configuration allows the basic ideas of the method to be presented without the complications that arise in more general cases. 

A monatomic gas is confined between the two parallel plates, which are kept at a constant temperature and an external acceleration is applied in the direction parallel to the plates. The magnitude of the external acceleration $F^\prime$ is sufficiently small to lead to a low-speed flow. 

The macroscopic description of this flow is based on the Cauchy momentum equation, which under the assumptions of a steady, fully-developed, one-dimensional, force-driven flow can be written as  
\begin{equation}
    \frac{dP_{xy}^\prime}{dy^\prime}=\rho^\prime F^\prime, 
    \label{eq:cauchy}
\end{equation}
where $P_{xy}^\prime$ is the shear stress and $\rho^\prime$ denotes the uniform gas density, with the superscript $(^\prime)$ denoting a dimensional quantity. It is important to note that the Cauchy momentum equation is always valid regardless of the Knudsen number. Introducing Newton’s law of viscosity 
\begin{equation}
    P_{xy}^\prime=-\mu_0^\prime\frac{du_x^\prime}{dy^\prime},
    \label{eq:newtonlaw}
\end{equation}
where $\mu_0^\prime$ is the gas viscosity at the reference temperature, to the Cauchy momentum equation, the Stokes equation is derived 
\begin{equation}
    \mu_0^\prime\frac{d^2u_x^\prime}{d{y^\prime}^2}=-\rho^\prime F^\prime.
    \label{eq:stokes}
\end{equation}
The Stokes equation is limited to the continuum and slip regimes, where Newton’s law is valid; in the transition and free-molecular regimes it fails to properly describe the gas behaviour. In order to facilitate the extension to large values of Kn, we utilise a corrected Newton’s law of the form
\begin{equation}
    P_{xy}^\prime=-\mu_0^\prime\frac{du_x^\prime}{dy^\prime}+\mu_0^\prime S^\prime\left(y^\prime\right),
    \label{eq:cornewtonlaw}
\end{equation}
where $S^\prime\left(y^\prime\right)$ is the correction that will be provided by a surrogate model. This corrected Newton’s law is introduced to the Cauchy momentum equation and the resulting equation can be written as
\begin{equation}
    \mu_0^\prime\frac{d^2u^\prime_x}{{dy^\prime}^2}=-\rho^\prime F^\prime+\mu_0^\prime\frac{dS^\prime}{dy^\prime}.
    \label{eq:macroeq}
\end{equation}
Equation (\ref{eq:macroeq}) is the basis of the macroscopic model in our MMS approach. For convenience, the following dimensionless quantities are introduced 
\begin{equation}
    y=\frac{y^\prime}{H^\prime},\ u_x=\frac{u_x^\prime}{\upsilon_0^\prime},\ P_{xy}=\frac{P_{xy}^\prime}{P_0^\prime},\ F=F^\prime\frac{H}{\upsilon^{\prime2}_0},\ S=S^\prime\frac{H}{\upsilon^\prime_0}.
    \label{eq:dimenquantities}
\end{equation}
Here, $H^\prime$ represents the plate separation distance, $P_0^\prime$ denotes the reference pressure, $\upsilon_0^\prime = \sqrt{2 k_B T_0^\prime/m}$ is the most probable molecular speed, where $k_B$ is the Boltzmann constant, $T_0^\prime$ denotes the reference temperature and $m$ is mass of the gas species. Introducing the dimensionless quantities of Eq.~(\ref{eq:dimenquantities}) to Eqs.~(\ref{eq:cornewtonlaw}) and~(\ref{eq:macroeq}), the dimensionless form of the corrected Newton's law
\begin{equation}
    P_{xy}=-\frac{2 \mathrm{Kn}}{\sqrt{\pi}}\frac{du_x}{dy}+\frac{2 \mathrm{Kn}}{\sqrt{\pi}}S(y),
    \label{eq:cornewtonlawDimen}
\end{equation}
and macroscopic equation
\begin{equation}
    \frac{d^2u_x}{dy^2}=-\frac{\sqrt\pi}{\mathrm{Kn}}F+\frac{dS}{dy},
    \label{eq:macroeqDimen}
\end{equation}
are obtained.

In typical continuum fluid mechanics, the no-slip boundary conditions are used, which for the current flow configuration can be written as
\begin{equation}
    u_x(y_{bot})=0,\ u_x(y_{top})=0,
    \label{eq:noslipbc}
\end{equation}
where $y_{bot}$ and $y_{top}$ indicate the $y-$coordinate of the bottom and top boundaries, respectively. Slightly rarefied gas flows can be modelled using the typical continuum equations with slip boundary conditions 
\begin{equation}
    u_x(y_{bot})=-\sigma_P P_{xy}(y_{bot}),\ u(y_{top})=\sigma_P P_{xy}(y_{top}).
    \label{eq:slipbc}
\end{equation}
In the slip regime, where the constitutive laws are still valid and the slip boundary conditions are typically used, theoretical values of the slip coefficient $\sigma_P$ can be obtained. In the present work, correlations between the velocity and the shear stress at the boundary in the form of Eq.\ (\ref{eq:slipbc}) are made, using an appropriate surrogate model, for all Knudsen numbers. It is important to note that these surrogate models are not general, and can only be reliably used when supported by the DSMC data. 

\subsection{Surrogate Modelling} \label{sec:surrogates}

In this work, we use surrogate models which are simple probabilistic expressions that provide corrections to CFD in the form of more accurate boundary conditions and constitutive relationships. To do so, DSMC data is utilised to train the surrogate models, i.e.\ to find the proper model parameters, using Bayesian Ridge Regression. Bayesian Inference is then used to select on-the-fly the appropriate model that best describes the DSMC data without overfitting.

The surrogate models for the boundary velocity and constitutive equations are presented next. In order to keep the notation uncluttered, the subscripts of the shear stress and velocity are dropped, i.e. $P\equiv P_{xy}$ and $u\equiv u_x$.

Two surrogate models of the boundary velocity are considered, namely the no-slip and first-order slip models. In the present notation, the no-slip surrogate model is written as
\begin{equation}
    u^{(\mathrm{DSMC})}_j(y_{bot})=\epsilon_j,\ u^{(\mathrm{DSMC})}_j(y_{top})=\epsilon_j,
    \label{eq:noslipmod}
\end{equation}
where $j$ is the DSMC time index, i.e. $t=j\Delta t$, and $\Delta t$ the DSMC time-step. It is assumed that the noise  $\left(\epsilon_j\right)$ follows a normal distribution with zero mean and precision (inverse variance) $\beta$, i.e. $\epsilon_j\sim \mathcal{N}(0,\beta^{-1})$. The first order viscous slip surrogate model can be written as
\begin{equation}
    u^{(\mathrm{DSMC})}_j(y_{bot})=-\sigma_P P_{j}^{(\mathrm{CFD})}(y_{bot})+\epsilon_j,\  \ u^{(\mathrm{DSMC})}_j(y_{top})=\sigma_P P_{j}^{(\mathrm{CFD})}(y_{top})+\epsilon_j,
    \label{eq:slipmod} 
\end{equation}
where $P_j^{(\mathrm{CFD})}$ is the shear stress provided by Eq. (\ref{eq:cornewtonlawDimen}) at the current time step. The slip coefficient $\sigma_P$ is inferred from the available DSMC data, instead of using the theoretical value, for generality. For more general flows, higher-order slip models, or models containing more driving forces, such as thermal slip, can be used.

A hierarchy of constitutive law surrogate models $C_k$ is constructed, written in a general form as
\begin{equation}
    E_{i,j}=P_{i,j}^{(\mathrm{DSMC})}+\frac{2 \mathrm{Kn}}{\sqrt\pi}\left.\frac{du^{(\mathrm{DSMC})}}{dy}\right\vert_{i,j}=\frac{2 \mathrm{Kn}}{\sqrt\pi}S_k(y_i)+\epsilon_{i,j},
    \label{eq:ckmod}
\end{equation}
where $E_{i,j}$ is the error of Newton's law at cell $i$ and time step $j$, and the subscript $k$ denotes the model order. A special case is the $C_0$ model with $S_0=0$, which corresponds to Newton's law. The stress correction term $\left( S_k \right)$ is expanded as a sum of model weights $a_l$ and basis functions that depend on the distance from each boundary. In this work we choose to utilise exponential basis functions, as Knudsen layers are broadly of this form \cite{lockerby2005}, meaning that the stress correction can be written as
\begin{equation}
	S_k(y)=\sum_{l=1}^{k}{a_le^{-b_l(y-y_{bot})}}-\sum_{l=1}^{k}{a_le^{-b_l\left(y_{top}-y\right)}},    
    \label{eq:stresscor}
\end{equation}
where the first term of the right hand side accounts for the influence of the bottom boundary and the second term for the influence of the top boundary. It is noted that the coefficients of model $a_l$ and the exponents of the basis functions $b_l$ are the same for both boundaries to enforce the symmetry of the flow, implicitly satisfying that $S_k(y-y_{bot})=-S_k(y_{top}-y)$. In the present work, linear supervised learning is used for model training, which restricts the surrogate models to be linear with respect to the model parameters. For this reason, the exponents $b_l$ in the stress correction, Eq. (\ref{eq:stresscor}) are given as input to the method. The values of the exponents for the basis functions used in this work were obtained by analyzing the results of the Knudsen layer for pressure-driven flow between parallel plates \cite{gibelli} and are given in Table \ref{tab:exponents}. These exponents increase in value as the Knudsen number becomes smaller to better capture the behaviour of the the non-equilibrium zone (Knudsen layer) that shrinks as $\rm Kn$ decreases. For more general flow configurations, where such information may not be available, a large collection of exponents and a sparse Bayesian learning method \cite{bishop, tipping, ard} could be used to determine which basis functions are needed.

\begin{table}
    \centering
    \begin{tabular}{c|ccccccc}
         $\rm Kn$&  0.01&  0.05&  0.1&  0.2&  0.5&  1& 5\\ \hline
         $b_1$&  130&  30&  20&  20&  10&  8& 7\\
         $b_2$&  150&  40&  40&  12&  7&  6& 6\\
         $b_3$&  200&  70&  10&  8&  6&  20& 5\\
    \end{tabular}
    \caption{Exponents of the basis functions for the stress correction.}
    \label{tab:exponents}
\end{table}

In order to facilitate training, the surrogate models are written as linear models of the form
\begin{equation}
    \mathbf{t}=\mathbf{\Phi} \cdot \mathbf{w}+\mathbf{\epsilon},
    \label{eq:generalmodel}
\end{equation}
where $\mathbf{t}$ is the target vector (i.e.\ the DSMC data to be fitted), $\mathbf{\Phi}$ is the design matrix of the model, $\mathbf{w}$ are the unknown weights or parameters of the model and $\mathbf{\epsilon}$ is the additive Gaussian noise~\cite{bishop}. As before, it is assumed that the additive Gaussian noise consists of independent random variables that follow identical normal distributions with zero mean and precision (inverse variance) $\beta$, i.e. $\mathbf{\epsilon}\sim\mathcal{N}(\mathbf{0},\beta^{-1}\mathbf{1})$, where $\mathbf{1}$ denotes the identity matrix. The size of the target vector equals the total number of data points $N$, the size of the weights vector is equal to the number of parameters of each model $M$ and the design matrix is therefore an $N\times M$ matrix with the matrix element $\Phi_{n,m}$ given by evaluating basis function $\phi_m$ on the input data vector $\mathbf{x}_n$., i.e.
\begin{equation}
    \Phi_{n,m} = \phi_m(\mathbf{x}_n),\;n=1\ldots N, m=1\ldots M.
\end{equation}
The design matrices along with the target and weights vectors for the models are given in \ref{sec:appendixA}. It is noted that the no-slip and $C_0$ (Newton's law) models, that offer no corrections to typical CFD, are zero order models $(M=0)$ and thus no basis function is associated with them. 

Bayesian Ridge Regression~\cite{bishop,tipping} is used for model training. The weights of each model are assumed to follow a zero-mean isotropic Gaussian prior distribution governed by a single precision parameter $\alpha$. The question becomes, what is the posterior distribution of the weights, given the DSMC data $\mathcal{D}$? This distribution is the one that maximises the evidence function (or marginal likelihood)~\cite{bishop, tipping, mackay}. The posterior distribution is characterised by the mean $\mathbf{m}$ and the covariance matrix $\mathbf{S}$. The value of the parameters that is used by the selected model is the mean of the posterior distribution. In the present work, model training using Bayesian Ridge Regression is performed using the scikit-learn library~\cite{scikit}. The implementation is based on the algorithm described in Appendix A of~\cite{tipping} where the hyperparameters are updated following~\cite{mackay}. 

Model comparison and selection is performed using Bayesian inference. During model training, the evidence function $p(\mathcal{D}|\mathcal{M}_i)$, which indicates the probability of each model $\left(\mathcal{M}_i\right)$ given the data $\left(\mathcal{D}\right)$ is calculated. The probability of a model given the observed data $\left( p(\mathcal{M}_i|\mathcal{D})\right)$ is obtained using Bayes' theorem
\begin{equation}
    p(\mathcal{M}_i|\mathcal{D}) = \frac{p(\mathcal{D}|\mathcal{M}_i)p(\mathcal{M}_i)}{\sum_{j=1}^M[p(\mathcal{D}|\mathcal{M}_j)p(\mathcal{M}_j)]},
    \label{eq:bayes}
\end{equation}
where the prior probabilities of all models $\left(p(\mathcal{M}_i)\right)$ are assumed to be equal. The selected surrogate models (boundary conditions and constitutive law) that are used by the macroscopic approach is the one with the highest probability. It is noted that surrogate models for the boundary conditions and constitutive law are selected independently. 

Surrogate modelling allows the calculation of the uncertainties of the results due to the uncertainties of the model weights. This calculation is performed using the Monte Carlo method \cite{mcm}. More specifically, when the models are selected, a number of trials are performed. In each trial, a set of values for the weights is sampled from their respective posterior distributions using the mean and covariance matrix calculated during training. The uncertainty in model parameters is then propagated through to output quantities by running the macroscopic model for each set of parameter values and calculating the standard deviation of the output quantities.

\section{Results and discussion}\label{sec:results}
\begin{figure}
    \centering
    \includegraphics[width=0.9\textwidth]{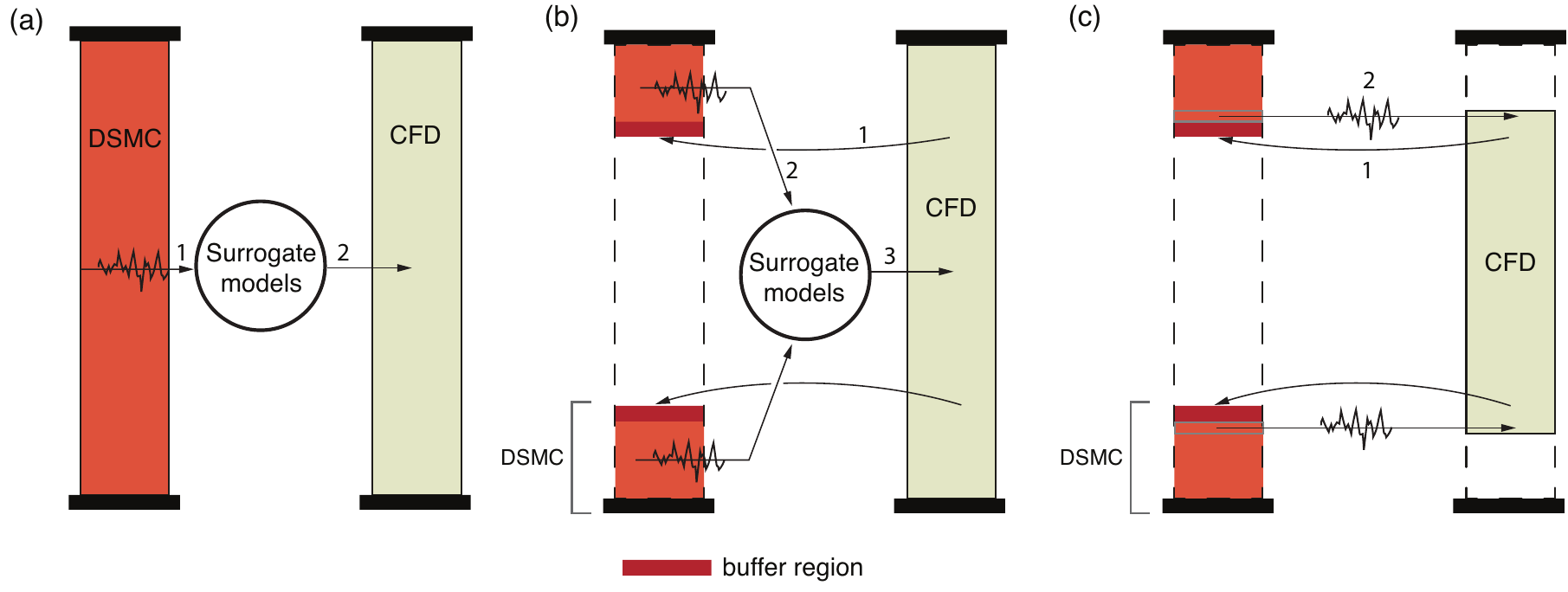}
    \caption{Different setups considered for the Poiseuille flow problem: (a) verification setup of the MMS surrogate coupling strategy (note: DSMC over full domain; no iterations); (b) the MMS hybrid (note: DSMC sub-domains in Knudsen layers overlaying CFD; iterations to steady-state) and (c) the conventional domain decomposition hybrid (note: direct coupling between DSMC and CFD; open boundaries for both).}
    \label{fig:coupling}
\end{figure}

Three cases for the one-dimensional Poiseuille flow system are presented in this section, as summarised in Figure \ref{fig:coupling}. The simplified setup, illustrated in Figure~\ref{fig:coupling}(a), that is used to verify the MMS surrogate coupling strategy is presented in Section~\ref{sec:1waycoupling}. Next, results based on the MMS hybrid, shown in Figure~\ref{fig:coupling}(b), are presented in Section~\ref{sec:2waycoupling} to verify the accuracy of the method. Section~\ref{sec:2waycoupling} also presents a comparison between the MMS hybrid and the direct (conventional) domain decomposition hybrid, shown in Figure~\ref{fig:coupling}(c), to highlight the advantages of the proposed approach.

In all cases, an external acceleration $F$ is applied to both the CFD and DSMC regions, and it is varied with the Knudsen number to generate a flow with Mach number around 0.05, calculated using the maximum flow velocity at the centre of the flow domain. The cell size for DSMC is set to $\Delta y=0.005$, and 50 particles per cell are used. All MMS-generated results are compared with a benchmark DSMC solution, that is obtained using 500 particles per cell and sampling for $10^6$ time steps after reaching the steady state to reduce the statistical noise.

For the MMS approach, the DSMC results are time averaged over 100 time steps to produce one training sample for the training of the surrogate models. Training is performed in batches, with each training step adding 100 training samples. In the present work, a fixed number of 1000 training steps is simulated, while in the future a criterion can be introduced to automatically detect when training should be terminated. The confidence intervals reported for the MMS results are two standard deviations from the mean, giving around a $95\%$ confidence interval, and are obtained by performing 1000 Monte Carlo trials.

\subsection{Verification of the MMS surrogate coupling}\label{sec:1waycoupling}

In order to demonstrate the accuracy of the MMS surrogate coupling strategy, we first present a case, illustrated in Figure~\ref{fig:coupling}(a), that is simple to verify. In this approach, both the microscopic and macroscopic models are applied across the whole flow domain, instead of through the constructed DSMC sub-domains. DSMC data is generated once, which is then used to train and select the appropriate boundary and constitutive surrogate models. These surrogate models are then used to correct the final macroscopic model solution. No iterative two-way coupling is employed.  

\begin{figure}
    \centering
    \begin{subfigure}[b]{0.49\textwidth}
        \includegraphics[width=\textwidth]{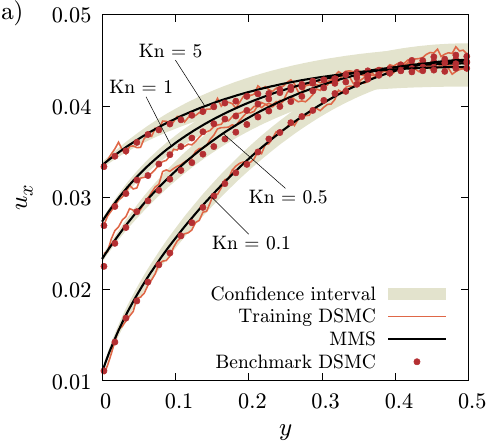}
    \end{subfigure}
    \begin{subfigure}[b]{0.49\textwidth}
        \includegraphics[width=\textwidth]{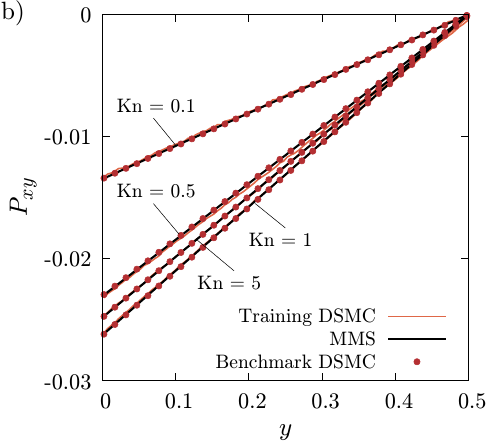}
    \end{subfigure}

    \caption{Velocity (left) and shear stress (right) distributions given by the: MMS approach along with confidence interval; DSMC used to train the surrogate models; benchmark DSMC, for various Knudsen numbers.}
    \label{fig:1wcoupl}
\end{figure}

Velocity profiles provided by the MMS approach along with their respective confidence intervals are presented in Figure~\ref{fig:1wcoupl}(a) for Knudsen numbers ranging from 0.1 to 5, covering the late transition regime. The MMS results are in excellent agreement with the benchmark DSMC solution. Moreover, the MMS solution is smooth in the flow domain, in contrast to the velocity distribution given by the DSMC used to train the surrogate models. The confidence interval is small close to the boundaries and increases towards the bulk of the flow. The boundary condition model is informed by less noisy data (velocity and shear stress) than the constitutive model (derivative of the velocity and shear stress) and thus the uncertainty associated with the boundary condition model is lower. As such, near the walls, where the solution is dominated by the boundary conditions, we are more confident in the prediction of the velocity profile than the bulk of the flow and the uncertainty quantification reflects this.

The respective comparison based on the shear stress profiles is shown in Figure~\ref{fig:1wcoupl}(b), where again an excellent agreement is observed. The confidence intervals for the MMS solution are not shown, as the macroscopic model always predicts the correct shear stress distribution, irrespective of the stress correction and slip coefficient, for this simple flow, as can be seen by combining Eqs.~(\ref{eq:cornewtonlaw}) and (\ref{eq:macroeq}).

\begin{figure}
    \centering
    \begin{subfigure}[b]{0.50\textwidth}
        \includegraphics[width=\textwidth]{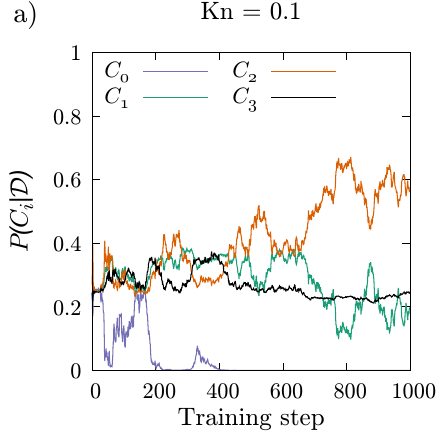}
    \end{subfigure}
    \begin{subfigure}[b]{0.48\textwidth}
        \includegraphics[width=\textwidth]{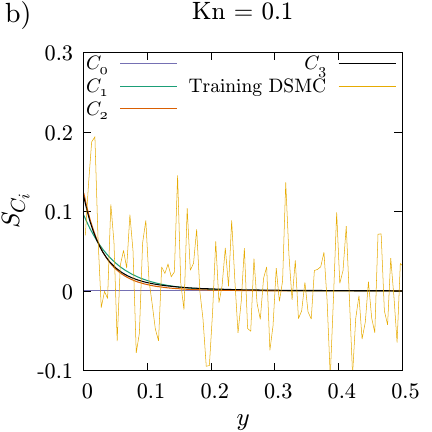}
    \end{subfigure}
    \begin{subfigure}[b]{0.50\textwidth}
        \includegraphics[width=\textwidth]{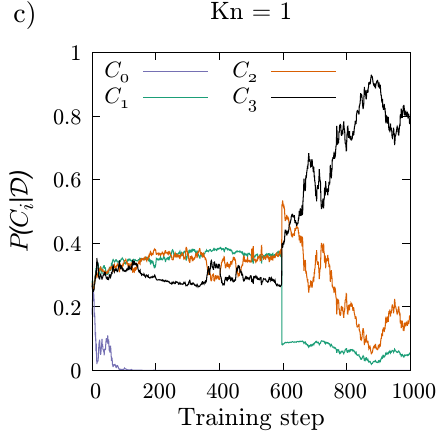}
    \end{subfigure}
    \begin{subfigure}[b]{0.48\textwidth}
        \includegraphics[width=\textwidth]{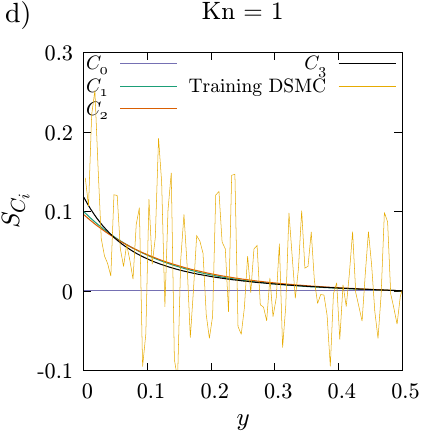}
    \end{subfigure}
    
    \caption{Evolution of the surrogate model selection probabilities for the  constitutive law with the number of training steps (left) and stress-correction profiles given by the different models at the final step along with the DSMC stress correction (right).}
    \label{fig:stress_models}{}
\end{figure}

Some observations are made about the surrogate models. For the boundary velocity, the slip model is always selected from the very first training steps with a  probability of model selection close to 1, as the data strongly suggest the existence of slip at the boundaries. However, further training steps are required to converge to the accurate value of the slip coefficient. Figure~\ref{fig:stress_models} shows the evolution of the probabilities of the constitutive law models with training steps and the final stress corrections given by each model, along with the stress correction calculated by using the raw training DSMC data, for $\rm Kn=0.1\ and\ 1$. Initially, all models have the same probability, as a uniform prior model probability is assumed. The $C_0$ (Newton's law) model is quickly discarded for the $\rm Kn=1$ case, while its probability is on par with the rest of the models for the $\rm Kn=0.1$ case, only for the first few steps. In the $\rm Kn=0.1$ case, the $C_2$ model is selected, while for $\rm Kn=1$, which is well within the transition regime, the more complex $C_3$ model is selected. It is observed that the stress corrections given by all models (except $C_0$) are very close in value, especially for small Knudsen numbers. A benefit of the Bayesian approach used for model training and selection, is that the simplest model explaining the data is naturally selected, i.e. Occam's razor is automatically applied, as can be seen especially for $\rm Kn=0.1$. It is worth mentioning, that the stress corrections are inferred from the very noisy training DSMC quantities shown. 

\subsection{Validation of the MMS hybrid method}\label{sec:2waycoupling}

The MMS approach can be used to model rarefied gas flows in a wide range of  Knudsen numbers, with accuracy comparable to DSMC, as demonstrated in the previous section. This section explores the potential of this framework to facilitate hybrid DSMC-CFD computations, where training and selection of the surrogate models occur only from the Knudsen layers. Results are presented for the MMS hybrid with DSMC sub-domains and with iterative coupling, as described in Section~\ref{sec:description} and shown in Figure~\ref{fig:coupling}(b). The MMS hybrid aspires to overcome some of the issues and limitations of the conventional direct coupling hybrid based on the domain decomposition, shown in Figure~\ref{fig:coupling}(c). A comparison between the two methods is also presented to demonstrate the effectiveness of the MMS approach.

For the direct hybrid simulations, the CFD covers only the middle part of the simulation domain, while DSMC covers the Knudsen layers, allowing for some overlap at the two open interfaces of the solvers, as expected in standard domain decomposition. In the present work, the DSMC subdomain size is two mean free paths from each boundary and the buffer region is 5 cells wide, for both hybrid methods. As in the MMS, the Chapman-Enskog distribution is applied at the interface of the DSMC, but now DSMC state quantities (right after the buffer region) are used as boundary conditions for the uncorrected typical CFD at its interface. The DSMC quantities are time averaged every iteration and the averaging is reset when the flow reaches steady state. In order to perform a fair comparison between the two hybrid methods, the direct hybrid uses the same DSMC parameters as the MMS hybrid and the same number of hybrid iterations and time steps per iteration are used for both hybrid methods.

\begin{figure}
    \centering
    \begin{subfigure}[b]{0.49\textwidth}
        \includegraphics[width=\textwidth]{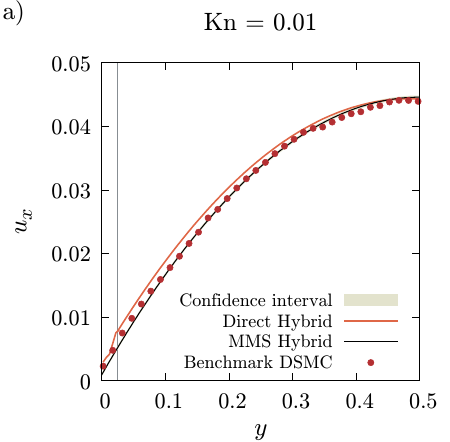}
    \end{subfigure}
    \begin{subfigure}[b]{0.49\textwidth}
        \includegraphics[width=\textwidth]{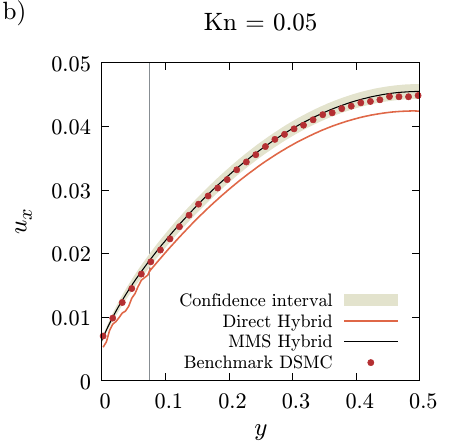}
    \end{subfigure}
    
    \vspace{0.3cm}  
    \begin{subfigure}[b]{0.49\textwidth}
        \includegraphics[width=\textwidth]{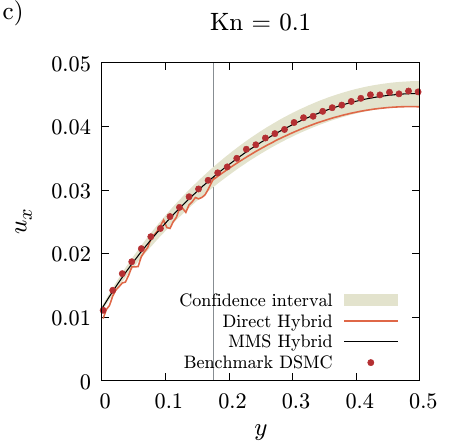}
    \end{subfigure}
    \begin{subfigure}[b]{0.49\textwidth}
        \includegraphics[width=\textwidth]{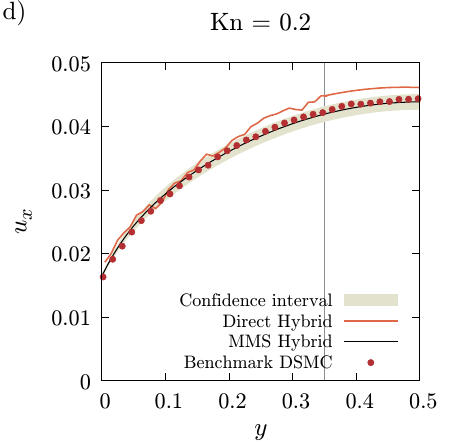}
    \end{subfigure}
    \caption{Velocity distributions for various Knudsen numbers given by the: MMS hybrid along with the confidence interval; direct coupling hybrid; benchmark DSMC. The vertical line indicates the interface between CFD and DSMC for the direct hybrid.}
    \label{fig:mms_hybrid_velocities}
\end{figure}

Velocity profiles are shown in Figure~\ref{fig:mms_hybrid_velocities} for $\rm Kn=0.01,0.05,0.1,0.2$ covering the slip to early transition regimes. The MMS hybrid results are shown along with their respective confidence intervals. Results based on the direct hybrid approach are also presented for comparison. The vertical lines indicate the location of the interface between CFD and DSMC for the direct hybrid approach, which coincides with the end of the training DSMC region for the MMS hybrid. The MMS hybrid results are in excellent agreement with the benchmark DSMC results. In the direct hybrid approach, when the same number of samples as the MMS hybrid is used, predictions can deviate significantly from the benchmark solution, indicating a lower level of accuracy in the overall solution. This is partly due to the fact that only the DSMC information of a single cell at the interface is used by CFD, and owing to the DSMC noise,  this leads to oscillating CFD solutions. In most cases, running the DSMC domains around 4 times longer in the direct hybrid method, allows the solution to be more stable, and closer to the MMS hybrid.

\begin{figure}
    \centering
    \begin{subfigure}[b]{0.495\textwidth}
        \includegraphics[width=\textwidth]{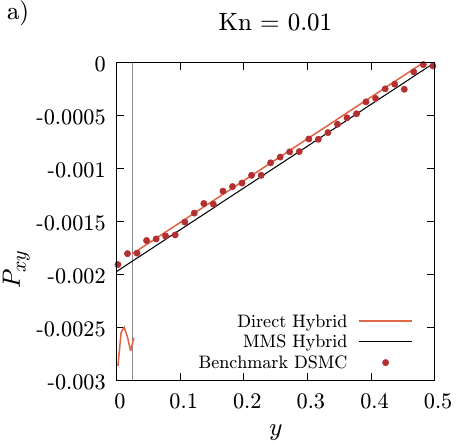}
    \end{subfigure}
    \begin{subfigure}[b]{0.49\textwidth}
        \includegraphics[width=\textwidth]{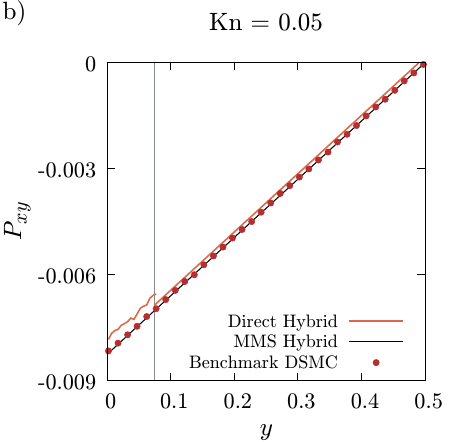}
    \end{subfigure}
    
    \vspace{0.3cm}    
    \begin{subfigure}[b]{0.49\textwidth}
        \includegraphics[width=\textwidth]{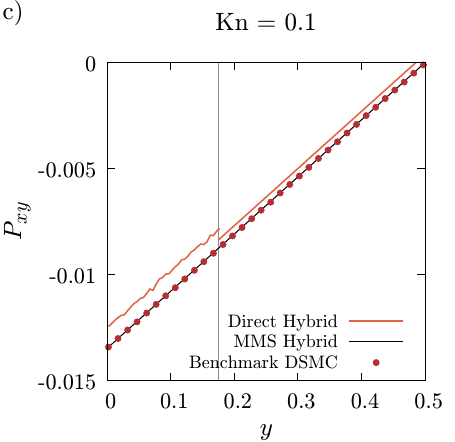}
    \end{subfigure}
    \begin{subfigure}[b]{0.49\textwidth}
        \includegraphics[width=\textwidth]{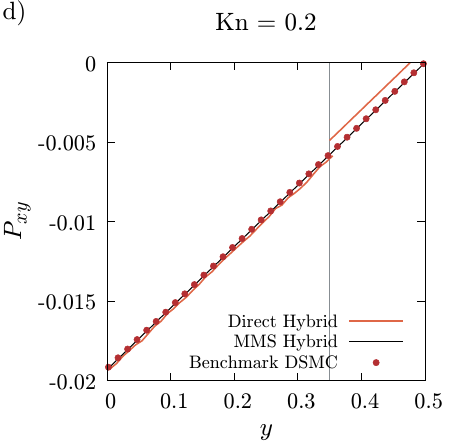}
    \end{subfigure}
    \caption{Shear stress distributions for various Knudsen numbers given by the: MMS hybrid; direct coupling hybrid; benchmark DSMC. The region between $y = 0$ and the vertical line indicates the DSMC micro region used in both direct and MMS hybrids.}
    \label{fig:mms_hybrid_stress}
\end{figure}

The corresponding comparison based on the shear stress is shown in Figure~\ref{fig:mms_hybrid_stress}. The MMS hybrid method is in excellent agreement with the benchmark DSMC. The direct hybrid on the other hand shows some deviations. For example for $\rm Kn=0.01$, the predicted shear stress in the DSMC regions is around 25\% larger than the actual value. The differences in the DSMC region for the direct hybrid can be attributed to the application of the Chapman-Enskog distribution as a DSMC boundary condition, which is informed by the local CFD quantities. 

\begin{figure}
    \centering
    \begin{subfigure}[b]{0.49\textwidth}
        \includegraphics[width=\textwidth]{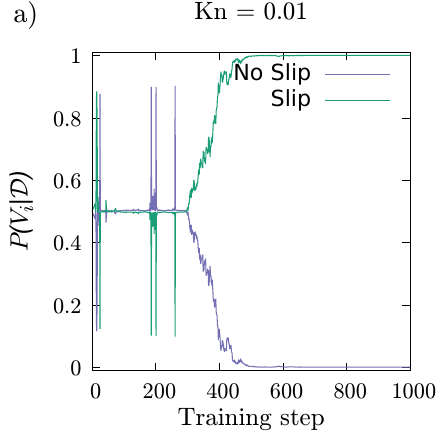}
    \end{subfigure}
    \begin{subfigure}[b]{0.49\textwidth}
        \includegraphics[width=\textwidth]{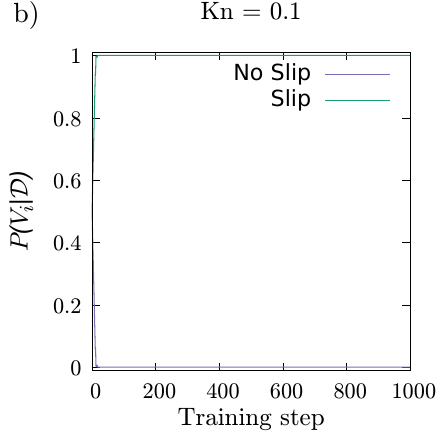}
    \end{subfigure}
    
    \caption{Evolution of the boundary condition model selection probabilities with the number of training steps for a) $\rm Kn = 0.01$ and b) $\rm Kn = 0.1$. }
    \label{fig:vel_models}{}
\end{figure}

The boundary condition model selection probabilities are shown in Figure~\ref{fig:vel_models} for $\rm Kn = 0.01$ and $\rm Kn = 0.1$. It is interesting to note that for the $\rm Kn=0.01$ case, the no-slip and slip models have roughly the same probability for a considerable part of the training period, as the slip velocity is rather small, although the slip model is ultimately selected. For $\rm Kn = 0.1$, where the slip at the boundaries is more pronounced, the slip model is selected from the very first training steps. 

It can be said that the MMS hybrid can outperform the direct hybrid in accuracy, when the same number of DSMC particles and timesteps are used across both setups. It should however be noted that the accuracy of the direct hybrid can be increased by increasing the number of DSMC particles and time steps, which may not always be possible especially for large problems.

\section{Conclusions}\label{sec:conclusion}
A novel micro-macro-surrogate (MMS) hybrid method that couples DSMC with CFD is presented for low-speed flows. The coupling of DSMC to CFD is based on surrogate modelling, where the DSMC data are used to train models for the boundary conditions and constitutive law, which in turn are used by CFD to provide an accurate noise-free solution. Model training and selection is performed using Bayesian inference. A verification of the surrogate coupling strategy is performed on a simplified case where the DSMC is applied in the whole computational domain. The MMS hybrid method is then formulated, where DSMC covers only the Knudsen layers. The MMS hybrid method provides an accurate and smooth solution, without the DSMC statistical noise, and has the additional computational cost benefit of typical hybrid methods.

In order to demonstrate the application of the method, a low-speed force-driven flow between parallel plates is simulated. The MMS hybrid is in excellent agreement with the DSMC benchmark. Additionally, it proves to be more accurate than the typical direct coupling hybrid method based on domain decomposition, as the destabilising DSMC noise is not passed on to CFD, but is filtered through the approximate physical models, which in turn inform macroscopic predictions. 

An added benefit of the MMS approach is that the uncertainty arising from micro-model fluctuations is automatically captured in the surrogate, as the parameters of each model are described by probability distributions. These distributions can then be used to quantify noise-related uncertainty in the overall predictions. 

To simulate more realistic cases, the method is being extended to include time-dependent flows and flows in higher dimensions. Moreover, additional physics can be included, such as temperature jump boundary conditions along with a correction to Fourier's law for flows with temperature differences. It is envisioned that once the appropriate extensions have been performed, the proposed MMS method can be used for multiscale simulations in cases where DSMC is too expensive to apply. 

\section*{Acknowledgements}
This work was supported by the Engineering and Physical Sciences Research Council (EPSRC), UK [grant numbers EP/V012002/1; EP/V01207X/1 \& EP/V012010/1].

\begin{appendices}
\section*{Appendix A: Surrogate model design matrices, weight and target vectors}
\label{sec:appendixA}
\setcounter{equation}{0}
\renewcommand{\theequation}{A.\arabic{equation}}
The appendix provides detailed information on the design matrices, weight and target vectors of the surrogate models used in this work.

\begin{itemize}[leftmargin=*]

\item Boundary condition models

The no-slip model, being a zero order model, does not have any model weight, basis function and data vector. For the slip model the data vector includes the shear stress at the boundaries, $\mathbf{x}_{2j-1}=-P_j^{CFD}(y_{bot}), \mathbf{x}_{2j}=P_j^{CFD}(y_{top}),$ while the basis function is $\phi(x)=x$. The design matrices, weight vector (slip model) and target vector for the boundary condition models can be written as:
\begin{itemize}\setlength{\itemsep}{10pt}
    \item No-slip model design matrix: 
    \begin{equation}
        {\Phi}^{(no-slip)}_{m}=0
    \end{equation}
    \item Slip model design matrix: 
    \begin{equation}
        {\Phi}^{(slip)}_{2j-1}=-P_{j}^{CFD}(y_{bot}), \quad \ 
    {\Phi}^{(slip)}_{2j}= P_{j}^{CFD}(y_{top})
    \end{equation}
    \item Slip model weight vector: 
    \begin{equation}
        \mathbf{w}^{(slip)}=\begin{bmatrix} \sigma_P \end{bmatrix}
    \end{equation}
    
    \item Velocity models target vector: 
    \begin{equation}
        {t}^{(V)}_{2j-1} = u_j^{DSMC}(y_{bot}), \quad \
        {t}^{(V)}_{2j} = u_j^{DSMC}(y_{top})
    \end{equation}
    
\end{itemize}
\item Constitutive law models

The $C_0$ model is a zero order model and as such it does not include any model weights, basis functions and data vector. For the more general models, the data vector is the cell centre coordinate, $\mathbf{x}_{i+(j-1)N_P}=y_i$. The basis functions used for the constitutive law models can be written as $\phi_k(y_i)=\frac{2\rm Kn}{\sqrt{\pi}}(e^{-b_k (y_i-y_{bot})}-e^{-b_k(y_{top}-y_i)})$. The design matrices, weight vectors and target vector for the constitutive law models can be written as:
\begin{itemize}
    \item $C_0$ model design matrix: 
    \begin{equation}
        {\Phi}^{(C_0)}_{i+(j-1)N_P}=0
    \end{equation}

    \item $C_1$ model design matrix: 
    \begin{equation}
        {\Phi}^{(C_1)}_{i+(j-1)N_P}=\frac{2 Kn}{\sqrt{\pi}}\left(e^{-b_1 (y_i-y_{bot})} - e^{-b_1(y_{top}-y_i)}\right)
    \end{equation}
    \item $C_2$ model design matrix: 
    \begin{equation}
        {\Phi}^{(C_2)}_{i+(j-1)N_P,l}=\frac{2 Kn}{\sqrt{\pi}}\left(e^{-b_l (y_i-y_{bot})} - e^{-b_l(y_{top}-y_i)}\right),\ l = 1,2
    \end{equation}

    \item $C_3$ model design matrix: 
    \begin{equation}
        {\Phi}^{(C_3)}_{i+(j-1)N_P,l}=\frac{2 Kn}{\sqrt{\pi}}\left(e^{-b_l (y_i-y_{bot})} - e^{-b_l(y_{top}-y_i)}\right),\ l=1,2,3
    \end{equation}
    \item $C_k$ models weight vectors: 
    \begin{equation}
        \mathbf{w}^{(C_1)} = \begin{bmatrix} a_1 \end{bmatrix},\
        \mathbf{w}^{(C_2)} = \begin{bmatrix} a_1\ a_2 \end{bmatrix}^T,\
        \mathbf{w}^{(C_3)} = \begin{bmatrix} a_1\ a_2\ a_3\end{bmatrix}^T
    \end{equation}
    \item Constitutive law models target vector: 
    \begin{equation}
        {t}^{(C)}_{i+(j-1)N_P}=E_{i,j}
    \end{equation}
    
\end{itemize}
\end{itemize}
In the above expressions, $m=1,\dots,2N_S$, $j=1,\dots,N_S$ with $N_S$ denoting the total number of training samples and $i=1,\dots,N_P$, with $N_P$ denoting the total number of points in the physical space between the two plates used for training.
\end{appendices}

\bibliography{biblio}
\end{document}